\documentclass[fleqn,usenatbib]{mnras}

\usepackage[T1]{fontenc}
\usepackage{ae,aecompl}

\usepackage{graphicx}	
\usepackage{amsmath}	
\usepackage{amssymb}	

\usepackage{graphicx}
\usepackage{alltt}
\usepackage{underscore}
\usepackage{url}
\usepackage[colorinlistoftodos]{todonotes}

\hypersetup{
 linkbordercolor={0 1 0}
}

\newcommand{\nebulos}{NebulOS}
\newcommand{\nbtt}[1]{\mbox{\texttt{#1}}}

\defcitealias{Btrfs}{Btrfs}
\defcitealias{Ceph}{CephFS}
\defcitealias{EC2}{EC2}
\defcitealias{Hadoop}{Apache Hadoop}
\defcitealias{Hama}{Apache Hama}
\defcitealias{HIPI}{HIPI}
\defcitealias{MapR}{MapR}
\defcitealias{Mesos1}{Apache Mesos}
\defcitealias{MPICH}{MPICH}
\defcitealias{Slurm}{Slurm}
\defcitealias{Spark}{Apache Spark}
\defcitealias{Storm}{Apache Storm}
\defcitealias{Torque}{TORQUE}
\defcitealias{Xen}{Xen}

\title[The \nebulos\ Framework]{\nebulos: A Big Data Framework for Astrophysics}

\author[N. R. Stickley \& M. A. Aragon-Calvo]{
Nathaniel R. Stickley,$^{1}$\thanks{E-mail: nstic001@ucr.edu}
Miguel A. Aragon-Calvo,$^{1}$
\\
$^{1}$Department of Physics and Astronomy, University of California, 900 University Ave, Riverside,
CA 92521 USA \\
}

\date{Accepted XXX. Received YYY; in original form ZZZ}

\pubyear{2016}

\begin{document}
\label{firstpage}
\pagerange{\pageref{firstpage}--\pageref{lastpage}}
\maketitle

\begin{abstract}
We introduce \nebulos, a Big Data platform that allows a cluster of Linux machines to be treated as
a single computer. With \nebulos, the process of writing a massively parallel program for a
datacenter is no more complicated than writing a Python script for a desktop computer.
The platform enables most pre-existing data analysis software to be used, as scale, in a datacenter
 without modification. The shallow learning curve and compatibility with existing software greatly
reduces the time required to develop distributed data analysis pipelines. The platform is built
upon industry-standard, open-source Big Data technologies, from which it inherits several
fault tolerance features. \nebulos\ enhances these technologies by adding an intuitive user
interface, automated task monitoring, and other usability features. We present a summary of the
architecture, provide usage examples, and discuss the system's performance scaling.
\end{abstract}



\section{Introduction}

In recent decades, the volume of data coming from experiments, sensors, observations, simulations,
and other sources has increased exponentially. Scientific disciplines that were once starved of data
are now being flooded with data that is not only massive in size, but heterogeneous and, in some
cases, highly interconnected. Consequently, scientific discoveries are increasingly being driven by
large-scale data analysis. Although many data analysis and management patterns are shared among
research groups, individual groups typically develop custom, application-specific, data analysis
pipelines. This results in duplicated efforts, wasted resources, and incompatibility among projects
that might otherwise complement one another.

In industry, the need to perform large-scale data analysis has resulted in the development and
adoption of data-aware frameworks, such as \citetalias{Hadoop}. Largely driven by Internet and
finance companies, these tools are most easily applied to Web and business data. Adapting these
tools for scientific data analysis is oftentimes not straightforward. Analogous tools, designed
specifically for large-scale scientific data analysis, management, and storage have not yet emerged,
despite the increasing need. In astronomy, for example, many projects, including the Sloan Digital
Sky Survey \citep{SDSS}, the Hubble Space Telescope, and the Bolshoi simulation \citep{bolshoi} have
each produced tens or hundreds of terabytes of data. Future projects, such as the LSST \citep{LSST},
will produce petabytes of data. While each of these projects produces different kinds of data and
performs different types of data analysis, the data management and analysis \textit{patterns} are
shared.

\subsection{I/O bottlenecks in conventional supercomputers}

Most large-scale scientific data analysis is currently performed on supercomputers in which
computing nodes are physically decoupled from data storage media. Such architectures are well-suited
for compute-intensive applications, such as large simulations, where CPU, memory, and inter-node
communication speed are the most important factors. However, they tend to perform poorly when
applied to data-intensive problems that require high data throughput, such as large-scale signal
processing, or analysing large ensembles of data. As the size of datasets approaches the petabyte
scale, traditional supercomputing architectures quickly become I/O-bound, which limits their
usefulness for data analysis.

\subsection{The Big Data approach}

A distributed computing architecture, known as the datacenter \citep{Hoelzle09}, has emerged as the
natural architecture for analysing the enormous quantities of data that have recently become
available. Like most supercomputers, datacenters consist of many machines connected via a network.
Unlike supercomputers, however, the data in most datacenters is stored close to the the computing
nodes. Typically, the data storage media (hard disk drives or solid state drives) are directly
attached to each computing node, which allows data to be transferred to CPUs at high speed without
passing over a network. This configuration topology is known as direct-attached storage (DAS).
Alternatively, a bank of hard drives can be located on each server rack. The computing nodes within
each rack then share the local bank of disks, using a local high-performance network. Depending on
the implementation details, this configuration is known as network-attached storage (NAS), a storage
area network (SAN), or a hybrid of NAS and SAN.

In order to make sure that each node in a datacenter primarily analyses data stored locally, rather
than needing to first transfer data over a cluster-scale network, special software frameworks have
been designed. These frameworks are commonly referred to as ``Big Data'' frameworks. Big Data
frameworks share three key features:

\begin{enumerate}
 \item They are aware of data placement. Most analysis is performed on machines which can read data
locally; instructions are executed on nodes that contain relevant data.

\item They are fault tolerant. Since datacenters often contain thousands of nodes, hardware failures
are frequent occurrences. Big Data tools automatically create redundant copies of the data stored in
the datacenter so that hardware failures do not result in data loss. Additionally, tasks that are
lost during a node failure are automaitcally rescheduled and performed on healthy nodes.

\item They provide an abstraction layer on top of the datacenter hardware. The typical user of a Big
Data framework does not need to be aware of the details of the underlying hardware.
\end{enumerate}

Big Data frameworks also often make use of new parallel programming models that efficiently use
datacenter hardware. The prime example of this is the MapReduce model \citep{MapReduce}, which
allows certain data analysis jobs to be automatically divided into many independent tasks and
performed in parallel on the nodes of the datacenter (the map step). The results of the map step are
then automatically combined into a final result (the reduce step).

Most existing Big Data frameworks require the user to have more software engineering expertise than
most astronomers possess. The language of choice for most of the frameworks is Java, which is
not a popular language among astronomers. These frameworks are also most easily used with
text-based data. Reading and writing binary file formats, common in scientific research, requires
extra effort and knowledge. Furthermore, the ability to use existing analysis software with these
frameworks is limited. Software usually needs to be aware of the framework in some way in order to
work properly. For a more detailed discussion of popular Big Data tools, refer to
Appendix~\ref{appendix2}.

\subsection{This work}

In this paper, we describe \nebulos\footnote[1]{\url{http://bitbucket.org/nebulos-project/}}, a Big
Data framework designed to allow a datacenter to be used in a seamless way by hiding the
complexities of the datacenter's hardware architecture from the user. \nebulos\ allows users to
launch any pre-existing command-line driven program on a datacenter; programs do not need to be
made aware of \nebulos, or the datacenter's architecture, in order to work. Any data format can be
read exactly as on a regular Linux-based workstation without extra effort. A Python module is
provided, allowing scientists to use the system interactively or from a script, and a C++ library
is also available.

In section~\ref{section:architecture}, we describe the general architectural ideas and specific
implementation details. Basic usage examples are included in section~\ref{section:examples}. We
discuss system performance in section~\ref{section:performance}. Finally, in
\ref{section:discussion}, we summarise \nebulos' strengths and limitations and discuss planned
work.


\section{\nebulos\ architecture}
\label{section:architecture}


The \nebulos\ platform consists of a distributed operating system kernel, a distributed file system,
and an application framework. The kernel and file system are industry-standard, open source
projects, managed by the Apache Software Foundation: \citetalias{Mesos1} \citep{Mesos2} and the
Hadoop Distributed File System (HDFS) \citep{HDFS}. The HDFS is mounted on each node of the system,
using FUSE-DFS (a tool included in the Hadoop software distribution), so that applications can
access the HDFS like a local file system. Using industry-standard tools for the core of the
platform allows us to focus our development efforts on the application framework.

The application framework provides a simple, intuitive interface for launching and managing programs
on the distributed system. The user only needs to be familiar with the basic usage of a Python
interpreter. No experience with distributed computing or multithreading is necessary. The framework
automatically schedules tasks so that, whenever possible, execution occurs on CPUs that have local
access to the data being used. This minimizes network traffic and maximizes the rate at which data
can be read. Tasks that are lost due to hardware problems, such as network interface card failure or
motherboard failure, are automatically rescheduled on other machines. The framework also provides
the user with a means of automatically monitoring each task that runs on the distributed system; a
user-defined task monitoring script can be executed at regular intervals to examine the detailed
behaviour of each task.

In the remainder of this section, we describe the three main components of the \nebulos platform.
The application framework is described in greatest detail because it is the component which makes
\nebulos\ unique.

\begin{figure}
\centering
\includegraphics[width=3.33in]{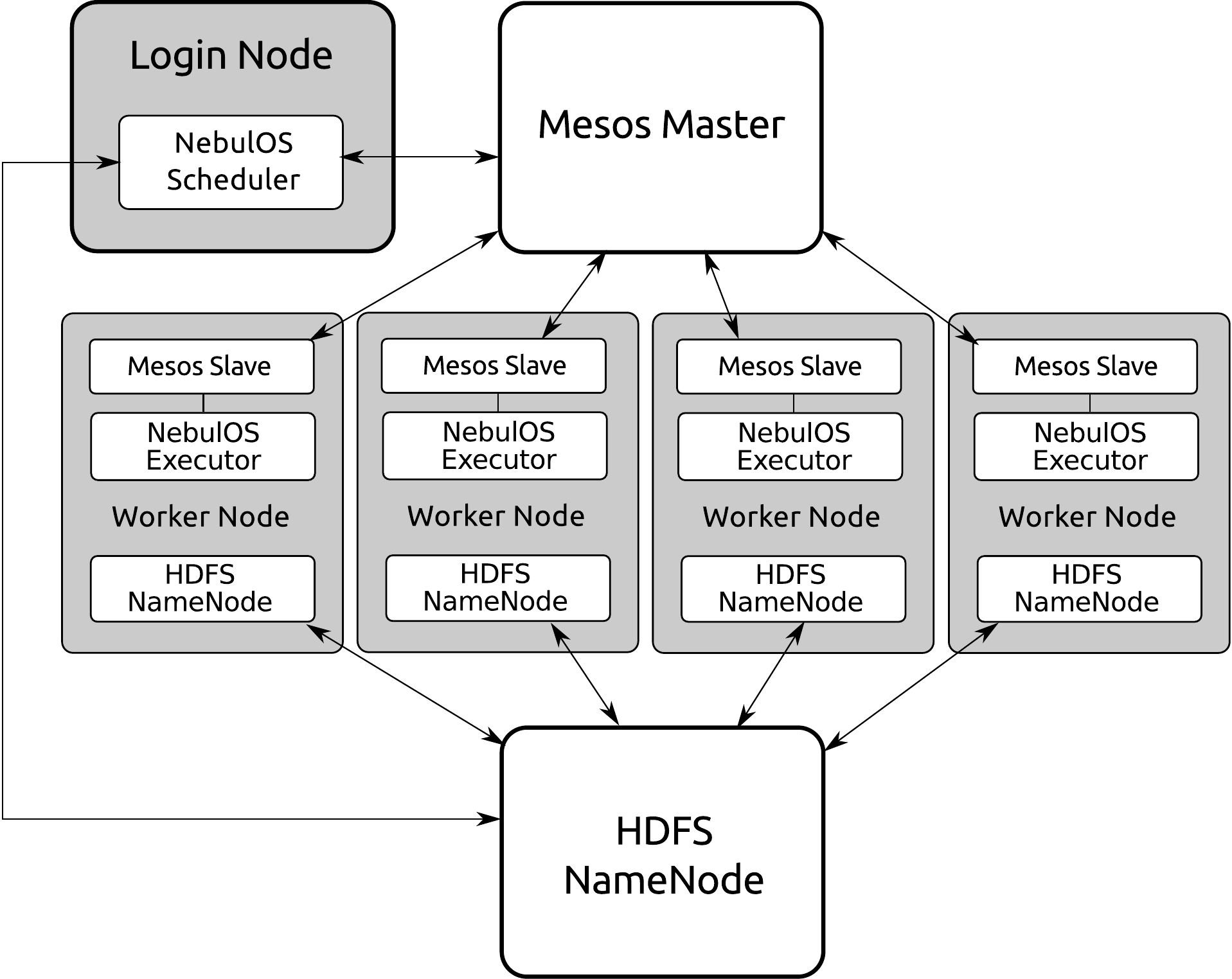}
\caption{An overview of the \nebulos\ architecture. Each worker node runs a Mesos Slave
daemon and an HDFS DataNode daemon. A Mesos Master daemon and HDFS NameNode daemon each
runs on a dedicated node. The scheduler of the \nebulos\ application framework runs on a
login node.}
\end{figure}

\subsection{Distributed resource management with Apache Mesos}

We chose Apache Mesos as \nebulos' distributed resource manager because of its scalability,
resiliency, and generality. Mesos is a distributed operating system kernel that manages resources on
a cluster of networked machines (e.g., a datacenter). Applications developed to run on top of Mesos
are called \textit{frameworks}. When the Mesos kernel detects that computational resources have
become available on the cluster, it offers these resources to a framework. The framework is then
responsible for selecting tasks to run on the resources offered by the kernel. A framework may also
decline resource offers that are not desired, in which case, the resources may be offered to another
framework. Mesos allocates resources to frameworks with a high degree of granularity; the smallest
allocatable resource unit is a single CPU thread. This makes it possible for tasks from multiple
frameworks to share a single machine. Consequently, Mesos allows the computer hardware in a
datacenter to be used efficiently, since there is no need to partition the datacenter into
application-specific sections.

\subsubsection{Mesos architecture}

Mesos consists of two components: a master daemon and a slave daemon. The master runs on a master
node of the system and an instance of the slave runs on each worker node. The master is responsible
for global resource management and system monitoring, while the slaves are responsible for managing
resources on individual nodes.

\begin{figure}
\centering
\includegraphics[width=3.33in]{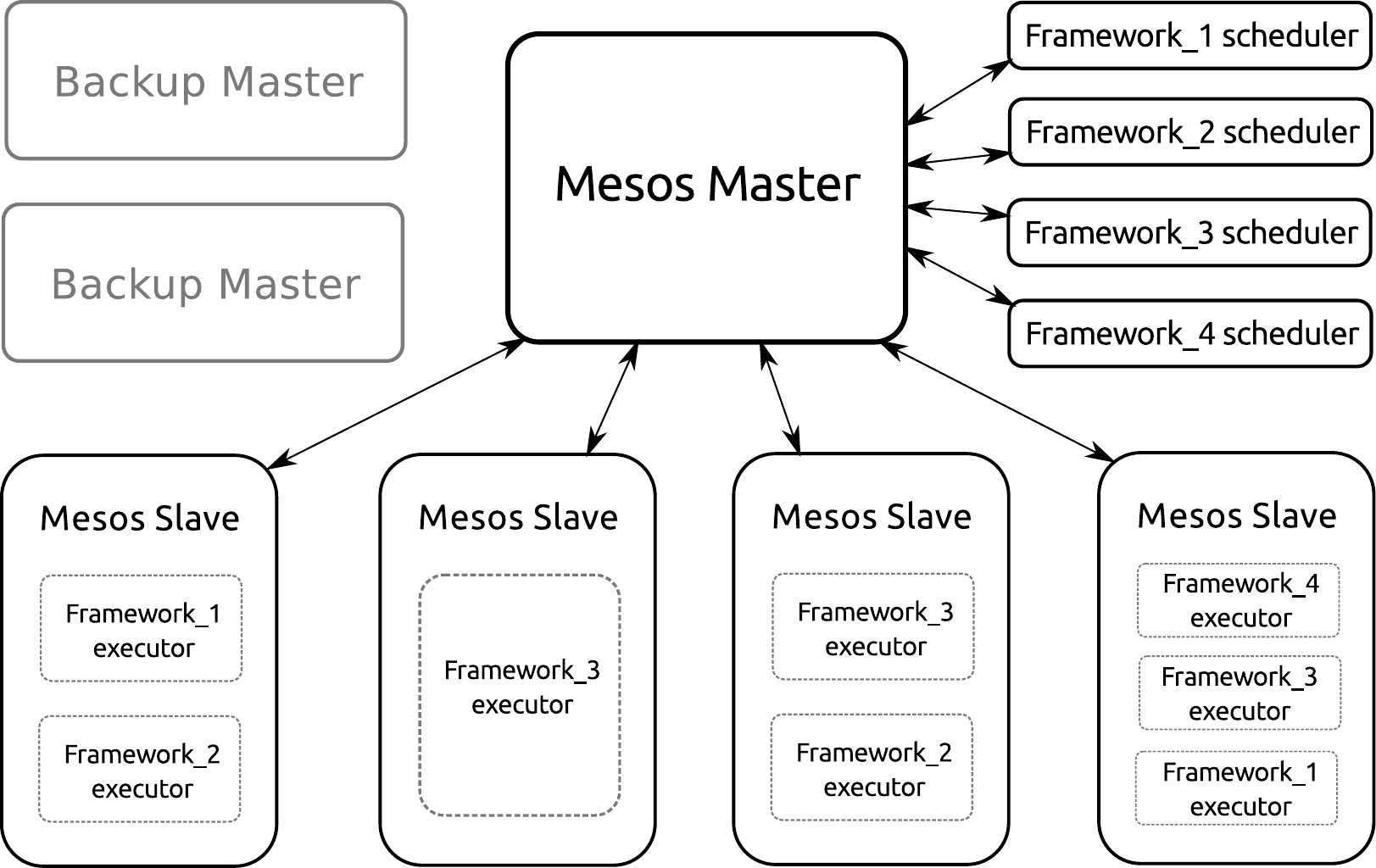}
\caption{\label{mesos}Mesos architecture overview. The Mesos slave daemon runs on each worker node
and communicates directly with framework executors. Note that multiple framework executors can share
a single slave node. Each slave daemon communicates with the Mesos Master node via a network.
Framework schedulers communicate with the Mesos Master. Optionally, the system can be configured so
that backup master nodes can take over for the active master node, in the case of hardware failure.}
\end{figure}

Each Mesos framework consists of two components, corresponding to the master-slave pair: a scheduler
and an executor. The scheduler handles resource offers and other information provided by the master,
such as task status update messages. The executors are responsible for performing the tasks assigned
to them by the scheduler and providing their local slave daemon with task status updates. Refer to
Figure~\ref{mesos} for a schematic overview of the Mesos architecture. For a discussion of Mesos'
fault tolerance features, refer to Appendix~\ref{appendix1}.

\subsection{Data locality awareness with HDFS}

HDFS is a robust, distributed file system, inspired by the Google File System architecture
\citep{GFS}. Files stored in HDFS are broken into blocks, which are then replicated on multiple
machines, so that the failure of any individual hard drive or host machine does not result in data
loss.

Suppose a user specifies that the block size for a particular file is 128 MB and that the block
replication factor for the file is three. If the file contains 500 MB of data, it would be broken
into four blocks and the HDFS would contain three copies of each block. The file could be
distributed across as many as 12 separate machines. When the file is later accessed, as many as 12
machines could potentially send data to the machine that is accessing the file. If \nebulos\ is used
to launch a program that reads this particular file, the program would automatically be launched on
the machine that contains the largest fraction of the file's data.

\subsubsection{HDFS architecture}

\begin{figure}
\centering
\includegraphics[width=3.33in]{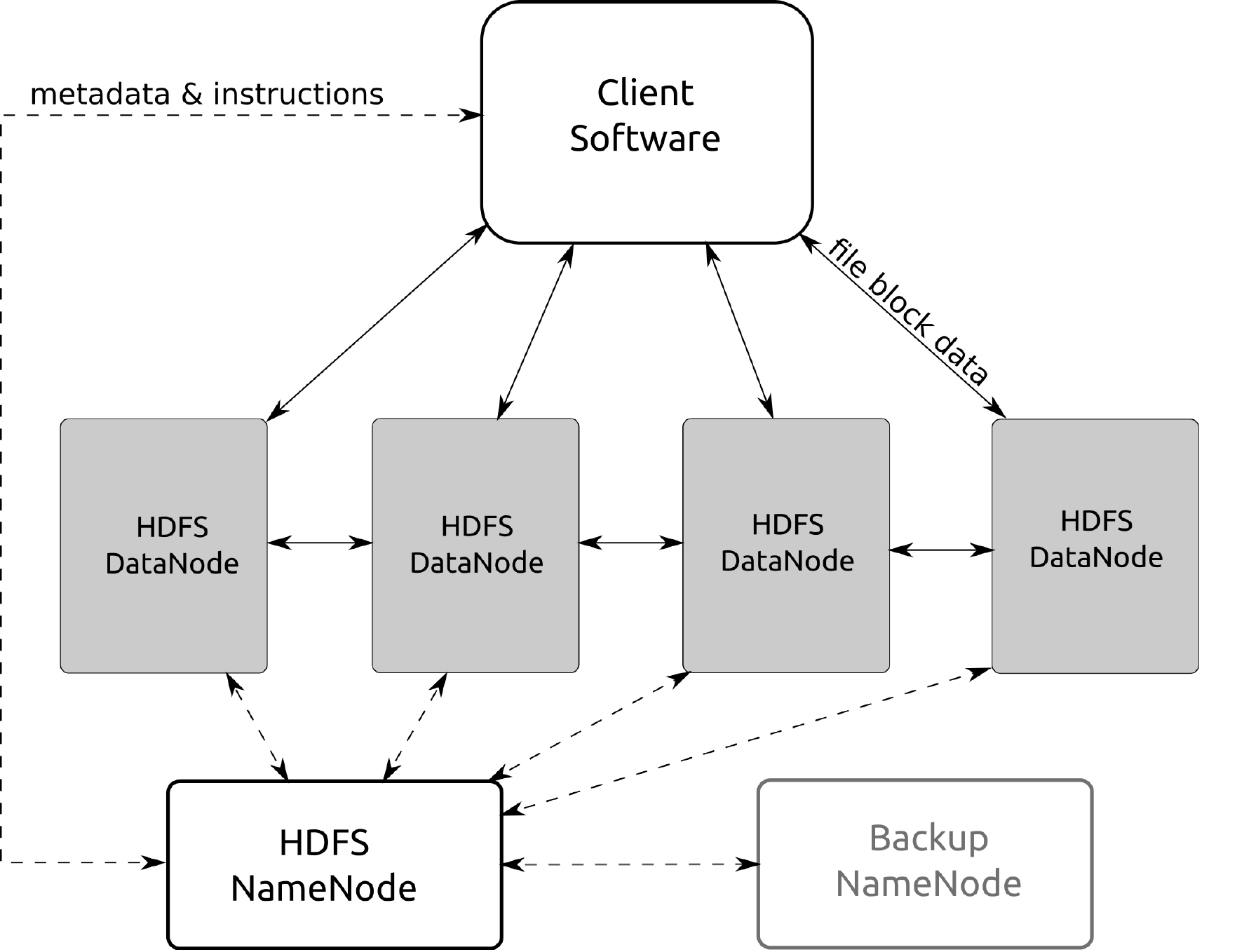}
\caption{\label{hdfs}HDFS architecture overview. Dashed lines indicate the transfer of metadata and
instructions, while solid lines indicate block data transfer. Client software sends instructions to,
and obtains metadata from, the NameNode. Each DataNode receives instructions and metadata from the
NameNode. File block data is transferred among DataNodes and between client software and DataNodes.
Optionally, a backup NameNode can be configured so that the system can survive a NameNode failure.}
\end{figure}

HDFS consists of two components: a NameNode daemon and a DataNode daemon. These are, in many ways,
analogous to Mesos' master and slave daemons. The NameNode stores the directory tree of the file
system, tracks the physical location of each block, and maps file names to file blocks. It also
ensures that each block is replicated as many times as specified by the user. Each DataNode daemon
manages blocks of data, which are stored on the local file system of the host machine. DataNodes are
responsible for performing actions requested by the NameNode. Data can be transferred between
DataNodes (in order to create copies) and directly between a DataNode and client software, as shown
in Figure~\ref{hdfs}.

To read data from the HDFS, client software first communicates with the NameNode, which provides the
identities of the DataNodes containing the blocks of interest. The client can then retrieve the
blocks of data directly from the relevant DataNodes. Transferring data to the HDFS proceeds
similarly; the client communicates with the NameNode to determine which DataNodes will contain
blocks of the file. The data is then transferred directly to the selected DataNode that is nearest
to the client, in terms of network distance. The nearest DataNode forwards packets of data to the
second-closest DataNode that was selected to contain a replica of the current block. This process
continues until the packet has been sent to all of the selected DataNodes.

We note that the HDFS is not a fully POSIX-compliant file system. For instance, once data has been
written to a file, it cannot be modified. Files can, however, be appended with new data and they can
be deleted and replaced with a new files with the same names as the old files.

\subsubsection{Standard file access with FUSE-DFS}

The HDFS must be accessed using a program that is aware of the HDFS interface. In order to allow
pre-existing software to access the HDFS without being modified, we use the FUSE-DFS utility, which
is part of the Hadoop software project. FUSE-DFS is used to mount HDFS as a local file system on
each node of the cluster. Any software can then access the HDFS as though it were an ordinary
directory on the local file system. Thus, the user does not need to modify their software in order
to take advantage of the features offered by \nebulos. However, FUSE-DFS imposes constraints on the
usage patterns. Most importantly, there is no support for appending data to a file. The user is not
presented with an error message when trying to append data to a file. Thus, the user must be careful
to not confuse the mounted HDFS with a regular file system.

\subsection{The \nebulos\ application framework}

\begin{figure}
\centering
\includegraphics[width=3.25in]{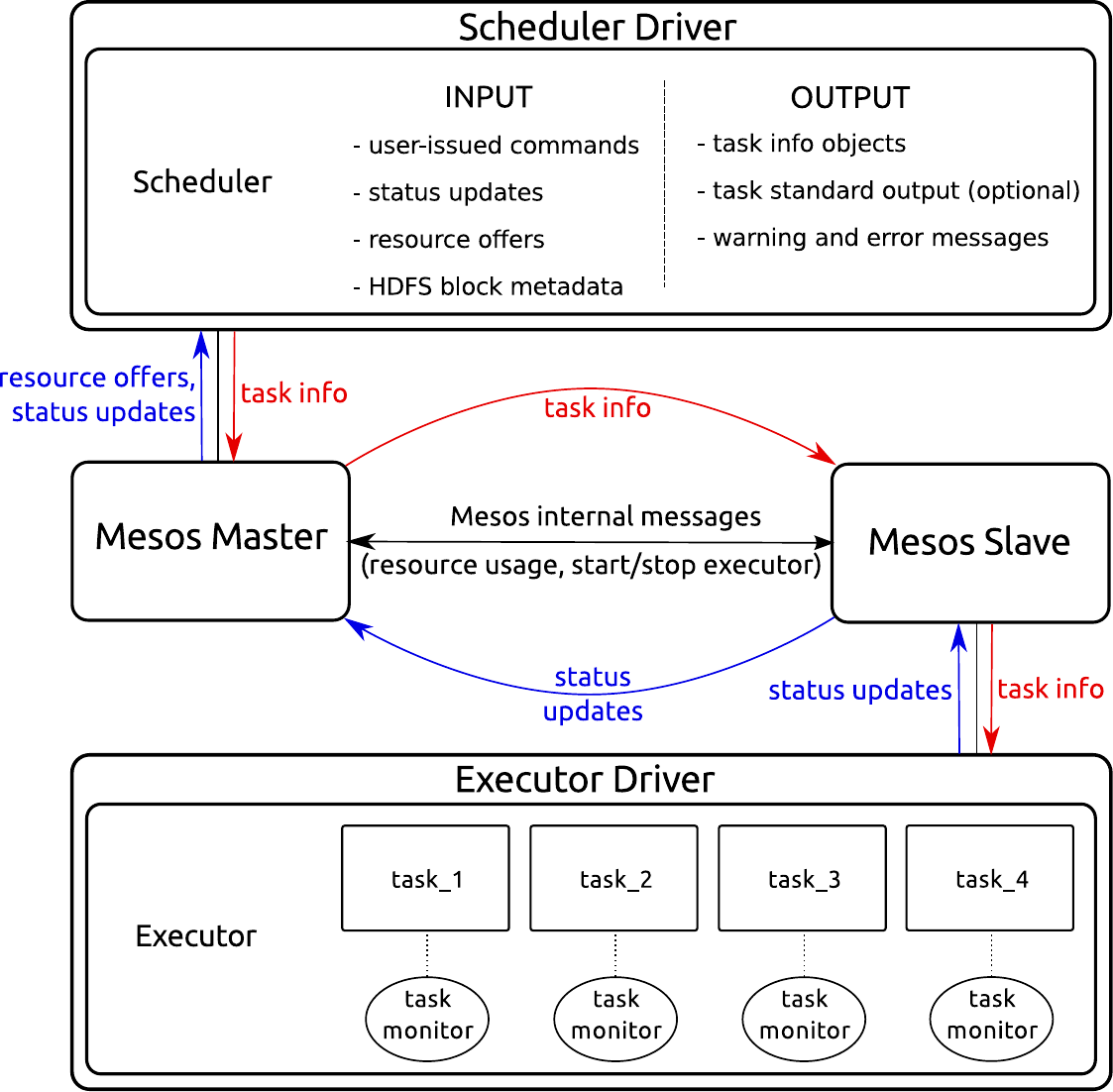}
\caption{\label{framework}An overview of the \nebulos\ application framework architecture. The Mesos
Master communicates with the Mesos Slave daemon, running on each worker node. The \nebulos\
scheduler interacts with the Mesos Master daemon, using the Mesos scheduler driver. Similarly, each
\nebulos\ executor interacts with its local Mesos Slave, using the Mesos executor driver. The
executors run tasks assigned by the scheduler and send status updates to the scheduler by way of the
Mesos Slave and Mesos Master daemons. Optionally, the executor can attach a task monitor to each 
task.}
\end{figure}

The \nebulos\ application framework is a custom Mesos framework that allows users to launch
arbitrary software on a cluster. The user simply specifies the commands that they wish to execute on
the cluster and the framework takes care of distributing and scheduling the tasks over the
individual computing nodes. A Python module and C++ library are available so that the framework can
be used interactively, via a script, or via a C++ application. The framework inherits the fault
tolerance offered by Mesos and HDFS and adds an extra layer of fault tolerance, in the form of task
monitoring scripts. Refer to Figure~\ref{framework} for a graphical summary of the communication
within the framework.

\subsubsection{The \nebulos\ schedulers}

The primary responsibility of the scheduler is to assign commands to appropriate host machines.
\nebulos\ provides two schedulers---a chronological scheduler, which simply launches tasks in the
same order in which they are provided by the user, and a DFS-aware scheduler, which schedules tasks
based on HDFS data placement. The operation of the chronological scheduler is trivial; whenever it
receives a resource offer, it accepts the offer and launches the next task in its internal FIFO
queue of tasks.

The DFS-aware scheduler is more complex. Each command assigned to the DFS-aware scheduler is first
inspected for the presence of HDFS file names. The scheduler then identifies which host machines
contain data used by each command. When the Mesos Master offers resources to the DFS-aware
scheduler, the scheduler checks to see which not-yet-launched tasks involve data on each host listed
in the offer. Tasks are then assigned to the appropriate hosts. If the scheduler is offered
resources on a host that does not contain any blocks of relevant data, the offer is declined.
Resource offers involving non-optimal hosts are only accepted if the scheduler has been unsuccessful
in requesting resources on a more suitable host. Typically, resource requests are only unsuccessful
if the appropriate hosts are being used by another framework. When the scheduler decides that a
non-optimal host must be used, it preferentially launches tasks which read the smallest amount of
data on these non-optimal hosts. This reduces network traffic and allows the scheduler to wait for
more desirable hosts to become available without wasting time.

\begin{figure}
\centering
\includegraphics[width=3.33in]{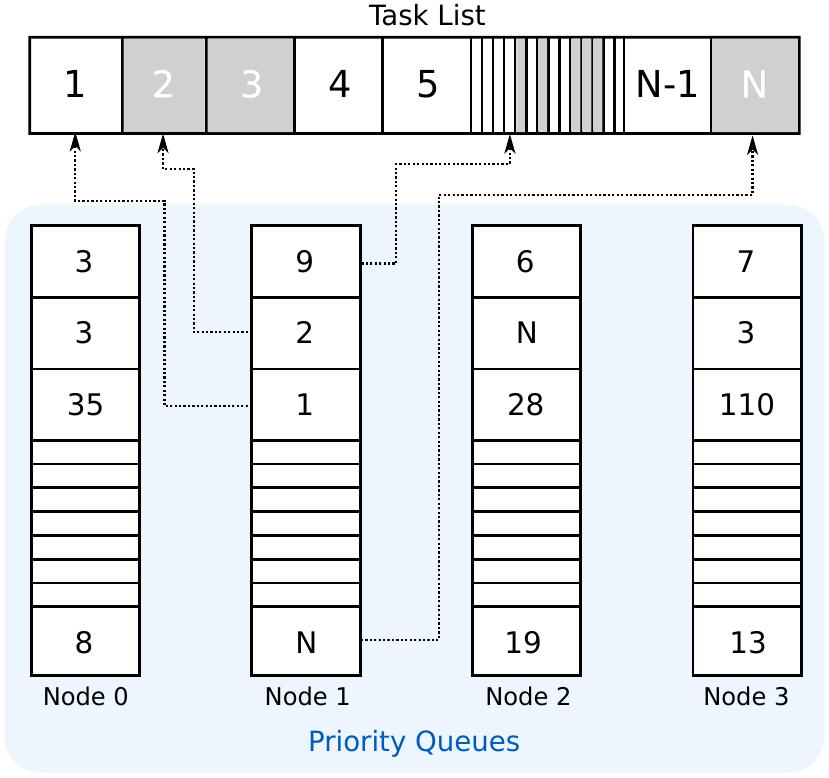}
\caption{\label{queues}Data structures used by the DFS-aware scheduler. All task information is
stored in an array, labled as ``Task List.'' Gray shading indicates tasks which have already been
launched, while white indicates that the tasks have not yet launched. For each worker node,
the scheduler creates a priority queue, which is populated with references to tasks that use data
stored on the associated node. The references in the priority queue for node Node~1 are shown
explicitly, as dotted lines. Suppose the Mesos Master offers resources for three tasks on Node~1.
The scheduler will check whether tasks 9, 2, and 1 have launched, by checking the task list. Since
task~2 has already launched, the scheduler will continue popping elements off of the queue until it
finds a task that has not been launched. Note that, during very long-running jobs, the scheduler
periodically re-builds the task list and stores information about completed tasks in a database on
the local disk.}
\end{figure}

Decisions regarding which task to launch next are made with the aid of priority queues, as
illustrated in Figure~\ref{queues}. For each worker node in the cluster, the DFS-aware scheduler
maintains a priority queue which returns task information based on the fraction of the task's data
that is located on the associated node. We call this fraction the \textit{residence fraction} of the
task. Tasks that involve reading files that are entirely stored on the host (i.e., tasks with a
residence fraction of 1.0) are of highest priority, while files with the smallest residence fraction
on the host are of lowest priority. This ensures that the scheduler efficiently takes advantage of
resources when they become available. When no optimal hosts are available, the sheduler launches
tasks which involve reading the smallest amount of data. In order to do this, the scheduler
maintains an additiona queue (not shown in Figure~\ref{queues}) that prioritizes tasks based on the
amount of data that they need to read; tasks which read the smallest amounts of data are of highest
priority in this queue. All of the priority queues are efficiently synchronized so that no tasks are
launched twice. Information about all unfinished tasks are stored in a list; the queues only contain
references to tasks in this list. When task information is returned from a priority queue, the
scheduler checks the status of the task. If the task has already been selected for launch, the
scheduler requests more tasks from the queue until it either encounters a task that has not been
previously selected or the queue is empty.

Whenever the status of a task changes, the Mesos Master informs the scheduler of the change by
sending a task status update message. The scheduler then makes a note of the change and takes
appropriate actions. For instance, if a task status message indicates that a task has been lost, the
scheduler assigns the task to a new host. If the status message indicates that a task has completed,
the content of the message is parsed for extra details. For instance, the message may contain
results of a computation performed by the task. The task status message may also indicate that a
particular task was killed by the executor. In this case, the message may contain a list of new
commands that should be launched on the cluster to replace the task that was killed.

\subsubsection{The \nebulos\ executor}

The executor is responsible for launching individual tasks on its host machine. It also sends status
update messages to the Mesos slave daemon, which then forwards the messages to the master. The
executor allows the standard output and error streams of each child task to be redirected.
For instance, the standard output can be saved to a file or stored in a status update
message. Note that the latter option is only practical for tasks that send a small amount of output
to the standard output stream. This is because status update messages are transferred to the master;
sending large amounts of data to the master would result in a communication bottleneck.

The executor can also execute a user-defined task monitoring script at regular intervals. The
monitoring script is provided with the content of the standard output and error streams and the
process identification number of the task that it is monitoring. If the script detects that the task
is behaving in an undesirable way, it can instruct the executor to terminate the task. Instructions
for re-starting terminated tasks can also be provided by defining a re-launch script. For example,
if the user knows that the phrase ``using defaults instead'' in a particular application's standard
error stream indicates that a non-fatal error has occurred, the user could write a script which
searches the standard error stream for that particular phrase. When this phrase is encountered, the
script can instruct the framework to terminate the task. If the user has defined a re-launch script,
the executor then runs this script immediately after terminating the task. The re-launch script can
be used to construct one or more new commands, which are then sent to the scheduler. For instance,
if the ``using defaults'' message depends on the input parameters of the code of interest, the
re-launch script could be designed to slightly alter the input parameters of the task that was
terminated.

\section{\nebulos\ use cases}
\label{section:examples}

In this section, we illustrate the basic usage of \nebulos\ by discussing a few examples.
All examples use the \nebulos\ Python interface.

\subsection*{Example I: Batch processing of images}

The most basic \nebulos\ use case involves using the framework as a batch processor. Suppose a large
number of images, stored in a datacenter, need to be processed using a program called
\nbtt{img_proc}, which requires two arguments: an output directory and an input filename. The
\nbtt{img_proc} program processes its input image, saves the resulting image to the specified output
directory, and writes statistics about the operation to the standard output stream. In order to use
\nbtt{img_proc} with \nebulos\ to process a directory full of images, the user places the
\nbtt{img_proc} program in the HDFS file system, opens a Python interpreter, and types the following
commands:

\vspace{3mm}
{\centering
\includegraphics[width=3.33in]{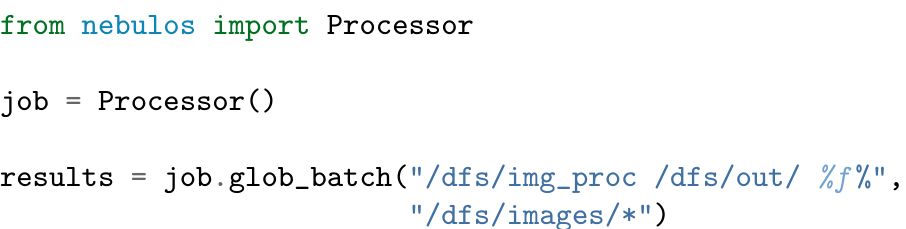}
}
\vspace{3mm}

\noindent
The first statement loads the \nebulos\ \texttt{Processor} class into the current namespace.
This class provides an interface to the \nebulos\ scheduler. Since the import statement is
always present, it will be omitted from subsequent examples.

The second statement creates a \texttt{Processor} object, named \texttt{job}, using default
parameters to set up a batch job. On the third line, the \texttt{Processor}'s \nbtt{glob_batch()}
method is used to construct a list of commands, which are then submitted to the \nebulos\ scheduler.
The second argument of \nbtt{glob_batch()} is a filename pattern, containing one or more wildcard
characters (asterisks). All files within the \nbtt{/dfs/images/} directory will be matched. The
first argument, which specifies the command to be launched, contains a file name placeholder,
\nbtt{\%f\%}. Suppose the directory, \nbtt{/dfs/images/} contains files named \nbtt{img_00000},
\nbtt{img_00001}, \nbtt{img_00002}, \ldots, \nbtt{img_50000}. In this case, the \nbtt{glob_batch()}
method would submit the following commands to the scheduler:

\begin{footnotesize}
\begin{alltt}
/dfs/img_proc /dfs/out/ img_00000
/dfs/img_proc /dfs/out/ img_00001
/dfs/img_proc /dfs/out/ img_00002
                \vdots
/dfs/img_proc /dfs/out/ img_50000
\end{alltt}
\end{footnotesize}

\noindent
By default, the standard output of each task is sent to the scheduler and is returned by the
\nbtt{glob_batch()} method as a Python list. Thus, \texttt{results} is a Python list containing the
standard output of each task.

Note that the placeholder \nbtt{\%f\%} indicates that the file is
located in the HDFS, so the DFS-aware scheduler will be used in this case. There is another
placeholder, \nbtt{\%c\%}, which can be used for arbitrary parameters. If the \nbtt{\%f\%} in this
example had been replaced with \nbtt{\%c\%}, the chronological scheduler would have been used,
instead of the DFS-aware scheduler.

\subsection*{Example II: Interactive data analysis}

\begin{figure}
\includegraphics[width=3.33in]{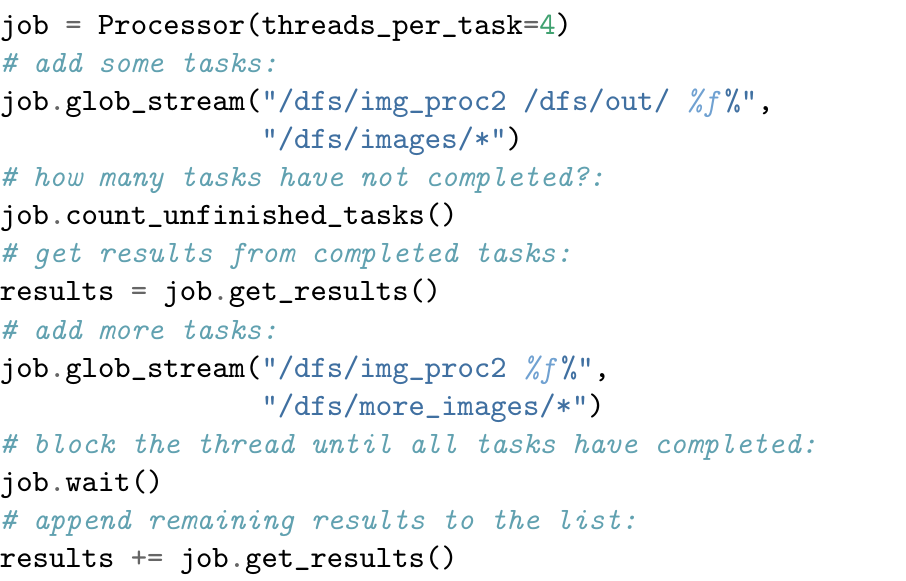}
\caption{Example~II \label{example2}}
\end{figure}

The \nbtt{glob_batch()} method, used in Example~I, causes the current thread to be blocked; the user
cannot work interactively with the batch job until all tasks have completed. In order to enable
interactive data analysis and task management, a streaming mode of operation is available. In this
mode, the user can inspect the status of each task, retrieve the output of completed tasks, add new
tasks to the scheduler, and cancel specific tasks before the entire job has finished. This type of
interactive usage is demonstrated in Example~II (Figure~\ref{example2}).

Example~II begins by providing the \texttt{Processor}'s constructor with a non-default parameter,
specifying that each task will be allocated four CPU threads. The \nbtt{glob_stream()} method on the
second line works the same way as the \nbtt{glob_batch()} method, discussed previously, except that
\nbtt{glob_stream()} does not block the thread. This makes it possible to retrieve a partial list of
results, add more tasks to the job, and perform various other operations while tasks are still
running on the cluster. The third command requests the number of tasks that have not yet finished
executing---a useful indicator of the job's progress. With the fourth command, the standard output
of each completed task is stored in a Python list, called \nbtt{results}; the standard output of the
tasks can be examined at this point. The fifth command assigns more tasks to the job. Calling
\nbtt{job.wait()} causes the thread to be blocked until all tasks have completed. Finally, we obtain
the remaining results by calling \nbtt{get_results()} once more. Note that individual results are
only returned \textit{once} by \nbtt{get_results()}. Thus, the second invocation of
\nbtt{get_results()} only returns the output from tasks that completed after the first invocation of
\nbtt{get_results()}.

\subsection*{Example III: Simple MapReduce implementation}

In streaming mode, multiple \nebulos\ schedulers can operate simultaneously and it is possible for
the schedulers to interact. In this example, a version of MapReduce is implemented using two
interacting schedulers.

\begin{figure}
\includegraphics[width=3.33in]{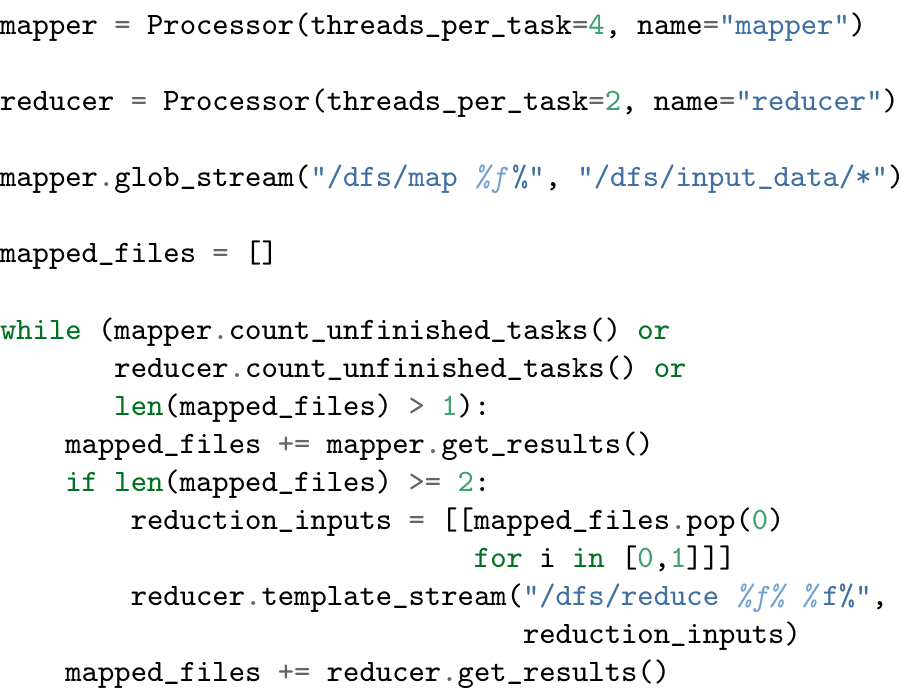}
\caption{Example III\label{example3}}
\vspace{-2mm}
\end{figure}

Suppose a program called \nbtt{map} analyses a single input file and saves the results of its
analysis in a new file whose file name is written to the standard output stream. Another program,
called \nbtt{reduce}, reads two files containing the output of the \nbtt{map} program. It then
summarizes the contents of the input files and saves the summary to a file whose name is written to
the standard output. These \nbtt{map} and \nbtt{reduce} programs are used together in Example~III
(Figure~\ref{example3}) to obtain a single file which summarizes the contents of a directory full of
input files.

The example begins with the creation of two \nbtt{Processor} objects, named \nbtt{mapper} and
\nbtt{reducer}, which will be used to run the \nbtt{map} and \nbtt{reduce} programs, respectively.
The \nbtt{mapper} tasks are allocated twice as many threads as the \nbtt{reducer} tasks because the
\nbtt{map} program requires more computational power than the reduce program. We have introduced the
\nbtt{name} parameter in the \nbtt{Processor} constructor, which allows us to easily distinguish
tasks belonging to different jobs. This allows us to easily distinguish between the \nbtt{mapper}
and \nbtt{reducer} if we need to inspect the Mesos logs.

The \nbtt{glob_stream()} method is then used to assign tasks to the \nbtt{mapper}. As tasks from the
\nbtt{mapper} are completed, the results are used to assign new tasks to the \nbtt{reducer}. Results
from the \nbtt{reducer} are recursively combined until only one output file name remains in the
\nbtt{mapped_files} list.

The \nbtt{template_stream()} method, used by the \nbtt{reducer}, constructs commands by substituting
instances of the parameter placeholders in its first argument with entries of the Python list in its
second argument. Multiple parameters can be specified by using a nested list as the second
parameter. Suppose the second argument is the following nested list:
\begin{small}
\begin{verbatim}
[["/dfs/temp/file_1", "/dfs/temp/file_2"],
 ["/dfs/temp/file_3", "/dfs/temp/file_4"],
 ["/dfs/temp/file_5", "/dfs/temp/file_6"]]
\end{verbatim}
\end{small}

\noindent
The \nbtt{template_stream()} method would submit the following commands to the
scheduler:

\begin{small}
\begin{verbatim}
/dfs/reduce /dfs/temp/file_1 /dfs/temp/file_2
/dfs/reduce /dfs/temp/file_3 /dfs/temp/file_4
/dfs/reduce /dfs/temp/file_5 /dfs/temp/file_6
\end{verbatim}
\end{small}

\noindent
Although it is not the most efficient implementation, this example hints at the ease with
which MapReduce can be implemented using \nebulos. Note that the \nbtt{mapper} and
\nbtt{reducer} streams are executed simultaneously and, because of the granularity offered by
Mesos, tasks belonging to the \nbtt{mapper} stream can be executed on the same host as tasks from
the \nbtt{reducer} stream.

\section{Performance characteristics}
\label{section:performance}

The performance achieved by \nebulos\ obviously depends upon the speed of the hardware on which it
is running. Thus, in our performance analysis, we focused on identifying how the performance of
\nebulos\ \textit{varies} as the number of files and worker nodes increases. In the following
analysis, we provide analytic estimates for the performance scaling, when possible, so that the
reader can estimate the performance of \nebulos\ on arbitrary hardware configurations.

\subsection{Methodology}
\label{section:methodology}

We were primarily interested in determining the speed with which data, already present in the HDFS,
could be analysed by software launched by \nebulos. We were also interested in the total overhead
time required for \nebulos\ to launch tasks and retrieve the standard output of the tasks. The data
reading speed, $R$, was simply determined using
\begin{equation}
  R = \frac{\rm dataset\, size}{\rm total\, elapsed\, time}, \label{eq:read-speed-definition}
\end{equation}
where the denominator is the total time that elapsed between submitting the commands to the
\nebulos\ framework and the arrival of the tasks' standard output at the Python interpreter. The
dataset consisted of a set of galaxy simulation snapshot files, produced by an $N$-body galaxy
simulation code. The files were dense binary files which did not benefit significantly from the
automatic file compression, employed by HDFS. The files were nealy uniform in size, with a mean size
of 110.8~MB and a dispersion of 0.4~MB. Each file in the dataset was analysed by a C++ program,
called \nbtt{reader}, which performed a trivial I/O bound task. Specifically, \nbtt{reader}
interpreted each as a list of 32-bit integers and counted how many of those integers were less
than $2\times10^6$. The performance of \nbtt{reader} was limited only by the speed of the data
storage medium from which the file was being read. The \nbtt{reader} program was installed on a
local hard drive of each node in the cluster so that it would not need to be read from the HDFS. A
list of snapshot file paths (file names) were stored in a Python list and we measured the time
required for the \nebulos\ framework's \nbtt{template_batch()} to complete. For example, to measure
the time required to read 1024 files, using the DFS-aware scheduler, we measured the total time
required for the following command,

\begin{footnotesize}
\begin{alltt}
output = job.template_batch("reader \%f\%", files[:1024])
\end{alltt}
\end{footnotesize}

\noindent
to complete, where the variable, \nbtt{files}, is a list containing all of our snapshot file names.

In order to determine the task-launching overhead, introduced by using Mesos and the \nebulos\
framework to launch tasks on the nodes of the cluster, we launched each task exactly as described
above, but, instead of actually reading the files, we only printed their names to standard output:

\begin{footnotesize}
\begin{alltt}
output = job.template_batch("echo \%f\%", files[:1024])
\end{alltt}
\end{footnotesize}

\noindent
where \nbtt{echo} is a standard Unix utility which prints its argument to the standard output
stream.

\subsection{Hardware configurations}

In our tests, we used two distinct hardware configurations, which we will refer to as ``MicroBlade''
and ``EC2.'' In both configurations, the roles of login mode, HDFS NameNode, and Mesos Master were
all performed by a single master node. The operating system used by all nodes was Ubuntu Linux
14.04.4, with the Linux 3.13-generic kernel. The HDFS block size was set to 128~MiB
($\approx134$~MB), which means that the files that we used during most of our tests were
single-block files.

\subsubsection*{MicroBlade cluster configuration}

The MicroBlade cluster consisted of eight worker nodes, running Mesos version 0.20.0 and HDFS
version 2.4.1. Each worker node was powered by a 4-core (8 simultaneous thread) Intel Xeon E3-1230
CPU, running at 3.3~GHz and containing 32~GiB of DDR3 RAM, running at 1,600~MHz. The hard drive
configuration was heterogeneous; all systems contained two hard drives, but half contained a total
of 7~TB of raw storage while others contained 5~TB. The HDFS block replication factor was set
to 2. Furthermore, the HDFS on the MicroBlade cluster was 89\% full.

The master node was powered by an Intel Core i7-4770K CPU running at 3.5~GHz with 32~GiB of DDR3
RAM, running at 2,133~MHz. The cluster's network interconnect was Gigabit Ethernet, with one active
Ethernet port per worker node (i.e., the HDFS and Mesos traffic shared the same Ethernet port).
The maximum network throughput achieved by each individual link in the network was $942\pm1$~$\rm
Mbit~s^{-1}$ (approximately 118~$\rm MB~s^{-1}$). The network ping time between the name node and
the worker nodes was $0.18\pm0.02$~ms. The operating system on all nodes was installed directly on
the hardware, rather than being installed in a virtual machine.

\subsubsection*{EC2 cluster configuration}

The EC2 clusters consisted of \citetalias{Xen}-based virtual machine instances running in
Amazon's Elastic Compute Cloud \citepalias{EC2}. Due to the way that the cloud service is designed,
some of the worker nodes potentially shared the same physical machines. The EC2 virtual machine
instances were placed into what Amazon Web Services refers to as a ``placement group,'' in order to
ensure that the network throughput and latency were optimal. The mean network ping time between
machines was $0.21\pm0.02$~ms and the mean network throughput between worker nodes was
$860\pm12$~$\rm Mbit~s^{-1}$. Mesos version 0.22.0 and HDFS version 2.6.0 were used for these
clusters. We used a block replication factor of 3 on the EC2 clusters. The only data stored in the
HDFS was our test dataset. Thus, the file systems were nearly empty.

The worker nodes used d2.xlarge EC2 instances, which were powered by 2 Xeon E5-2676 CPU cores (with
4 simultaneous threads) running at 2.4~GHz with 30.5~GiB of RAM. Amazon Web Services did not
disclose the exact type of RAM used by their instances, but the measured memory throughput of the
EC2 systems was approximately 88\% as high as that measured in the MicroBlade worker nodes. Each
worker node contained 3 hard drives for data storage---each with a capacity of 2~TB. The data drives
on each node were combined into a RAID-0 array, using the software RAID functionality built into
the B-tree file system \citepalias{Btrfs}. RAID-0 was used because it improved the performance when
reading individual HDFS blocks.

For the master node, we used a c3.8xlarge instance, which was powered by 32 Xeon E5-2680 cores
(64 simultaneous threads), running at 2.80~GHz. The master contained 60~GiB of RAM and its network
interface was capable of approximately 10~$\rm Gbit~s^{-1}$.


\subsection{Chronological versus DFS-aware schedulers}

\begin{figure}
\centering
\includegraphics[width=3.33in]{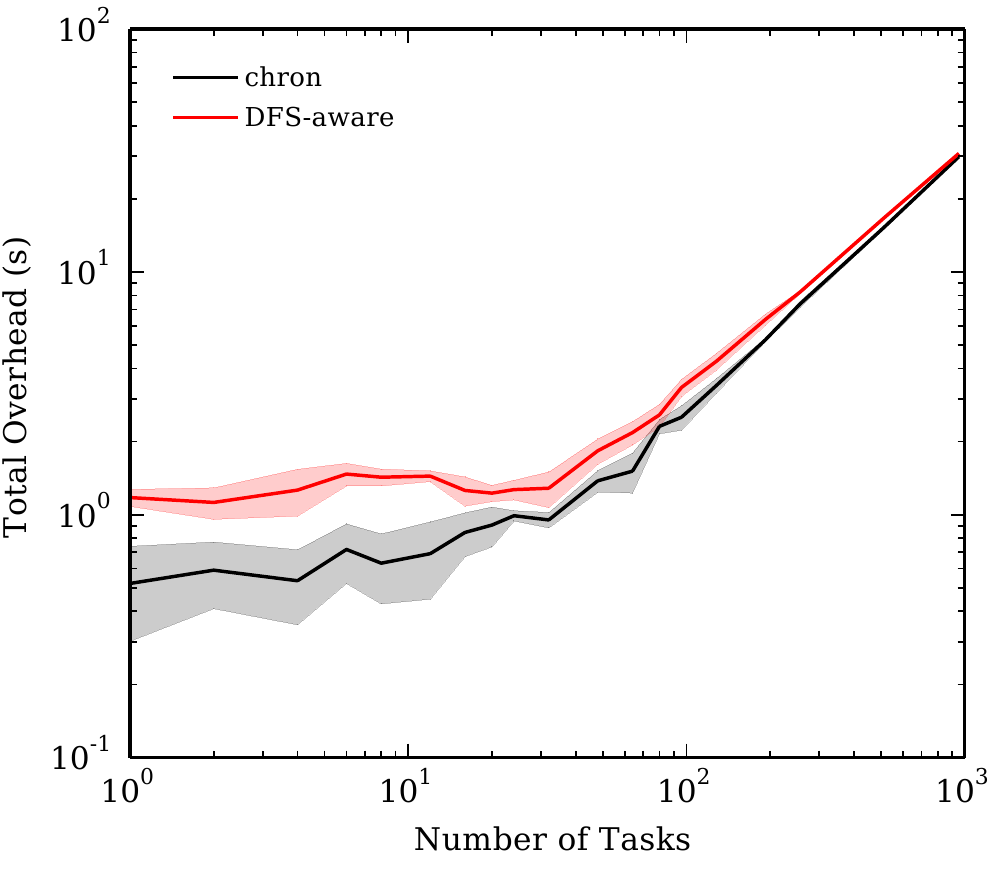}
\caption{\label{fig:scheduler-overhead}The total batch launching overhead on the MicroBlade system
with 8 worker nodes. The DFS-aware scheduler (red curve) requires more overhead than the
chronological scheduler (black curve), due primarily to the fact that the DFS-aware scheduler
communicates with the HDFS NameNode before launching tasks. Shaded regions indicate the 1-$\sigma$
scatter in overhead time.}
\end{figure}

Recall that the chronological scheduler launches tasks in the same order in which they are submitted
to the \nebulos\ framework; tasks are launched as quickly as possible, but the physical location of
the data being read by the tasks is ignored. On the other hand, the DFS-aware scheduler takes data
locality into account when launching tasks, so that tasks can be launched on machines that contain
the data. Figure~\ref{fig:scheduler-overhead} shows a comparison of the overhead required by these
two schedulers on a cluster with 8 worker nodes, when allocating 4 simultaneous tasks per node. The
DFS-aware scheduler requires more time to launch tasks because it first has to request file location
data from the HDFS NameNode.

Although the DFS-aware scheduler requires more overhead, it leads to a higher data reading speed
once more than a handful of tasks are being launched because it allows task software to read most of
the required data directly from disk, rather than transferring the data over the network. This is
demonstrated in Figure~\ref{fig:file-scaling}. The performance of the DFS-aware scheduler also
tends to be more consistent than that of the chronological scheduler. Note that the DFS-aware
scheduler can also take advantage of faster hard drives (or additional hard drives) more
effectively than the chronological scheduler, which is primarily limited by network throughput.

\subsubsection*{Data caching}

\begin{figure}
\centering
\includegraphics[width=3.33in]{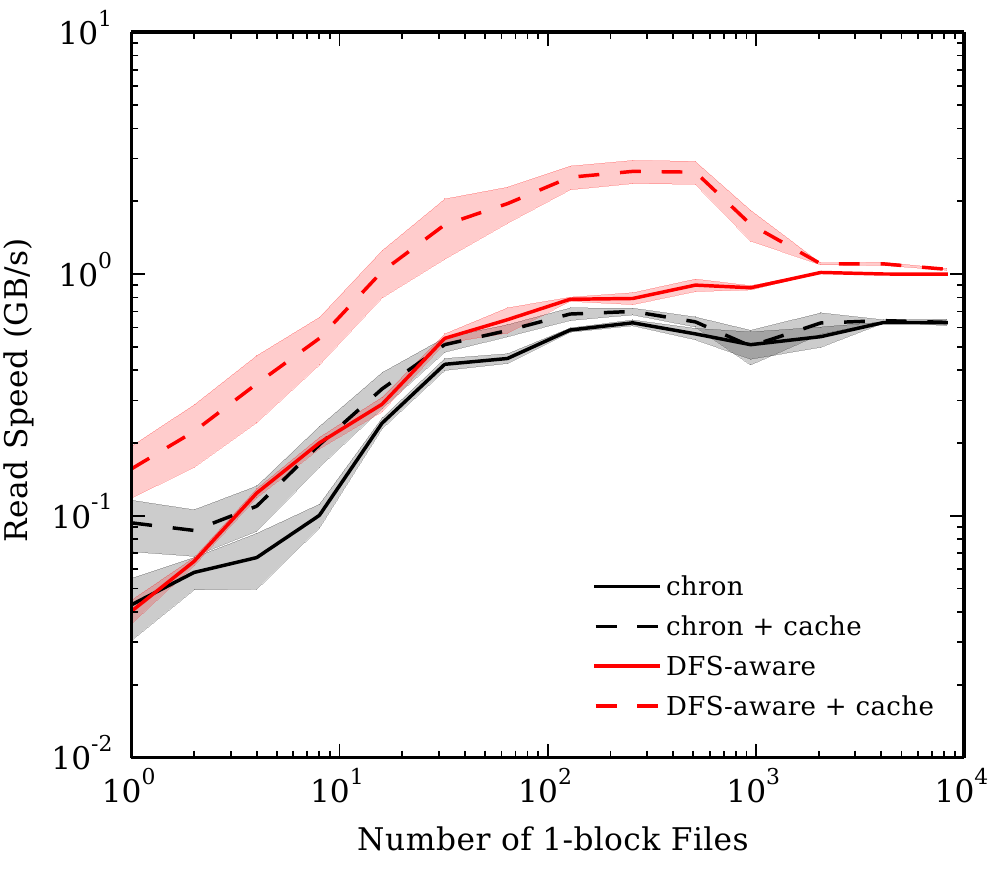}
\caption{\label{fig:file-scaling} A comparison of the read speed offered by the DFS-aware (red
curves) and chronological (black curves) schedulers on the MicroBlade system. Solid lines indicate
the performance when the caches are clean and the files are being read for the first time. Dashed
lines indicate the performance when the same files are read 6 times without dropping the Linux
kernel's automatic caching between trials. The shaded regions indicate the 1-$\sigma$ scatter in the
reading speed. Note that the DFS-aware scheduler generally outperforms the chronological scheduler.
Also note that the benefit of automatic caching diminishes once the volume of data being read
exceeds the size of the cache.}
\end{figure}

The current version of \nebulos\ does not have a built-in mechanism for caching data in memory for
faster performance when the same files need to be read repeatedly. However, since the Linux kernel
automatically caches data that is read from disks, \nebulos\ is able to read at a somewhat faster
rate when a set of files are repeatedly accessed. The DFS-aware scheduler is able to take advantage
of the Linux kernel's automatic caching more effectively than the chronological scheduler because
individual nodes are more likely to read the same files repeatedly when the DFS-aware scheduler is
used. This performance advantage is clearly visible in Figure~\ref{fig:file-scaling}. Once the
amount of data being read exceeds the cache size, the advantage of caching vanishes.

\subsection{Varied cluster size}

The main point of building a cluster for data analysis is to improve the speed with which data can
be analysed. Ideally, the speed of the cluster should scale linearly as the number of computers
increases; $N$ identical computers should be able to process data $N$ times faster than a single
computer. This ideal is typically not achieved, however. Even in the case of an embarrassingly
parallel data processing task, there is typically some sort of overhead involved in setting up a
parallel processing task on a cluster. For instance, the data first needs to be divided up among
the individual computers in a cluster. Then the data and instructions need to be sent to the nodes,
via a network, and the results of the tasks need to be collected. In this section, we explore
how \nebulos\ scales as the number of worker nodes increases. Since we know that the DFS-aware
scheduler performs better than the chronological scheduler, the analysis in this section only
involves the DFS-aware scheduler.

\subsubsection{Overhead}

\begin{figure}
\centering
\includegraphics[width=3.33in]{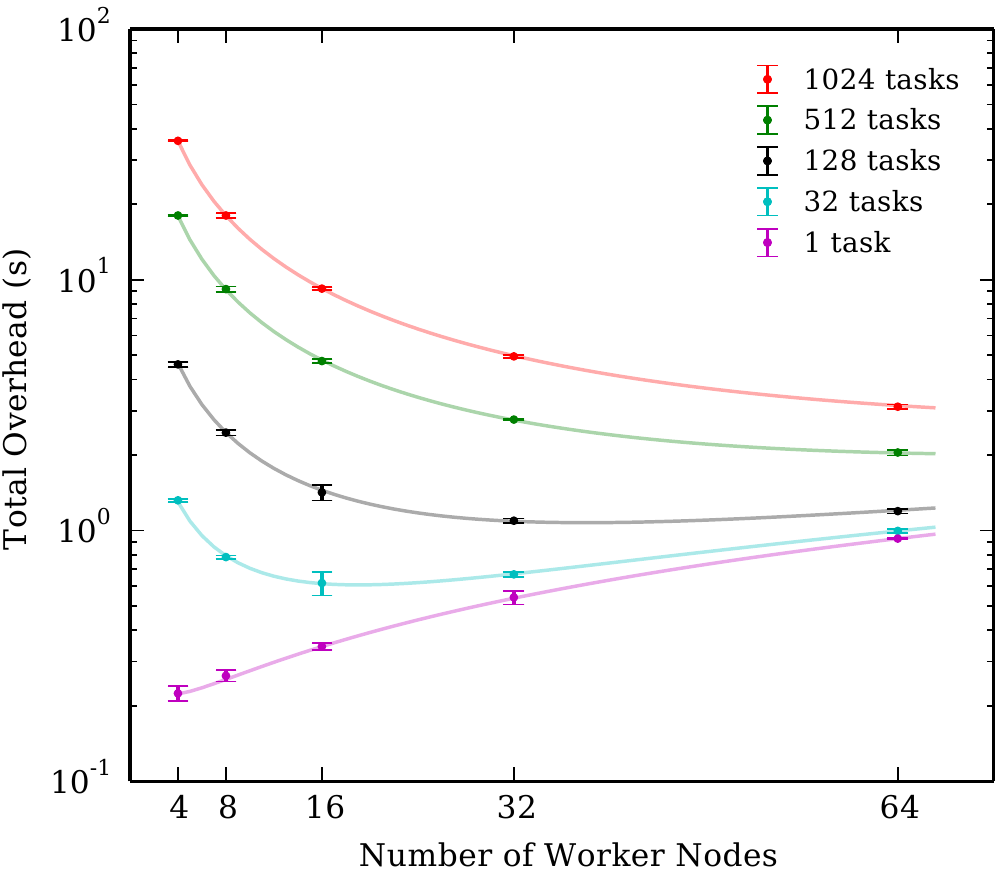}
\caption{\label{fig:node-overhead} The total overhead time as a function of cluster size when
launching various numbers of tasks. The error bars show actual measurements using the EC2 hardware
configuration with four simultaneous tasks per node (i.e., $\nu=4$), while the curves show the
overhead estimated using Eq.~\ref{eq:overhead}, with appropriately-tuned time constants. The
overhead for launching a small number of tasks increases as the cluster grows, but for large numbers
of tasks, the overhead decreases as the cluster grows. Here, ``small'' and ``large'' are relative
to the number of tasks that can simultaneously run on the cluster.}
\end{figure}

\begin{figure}
\centering
\includegraphics[width=3.33in]{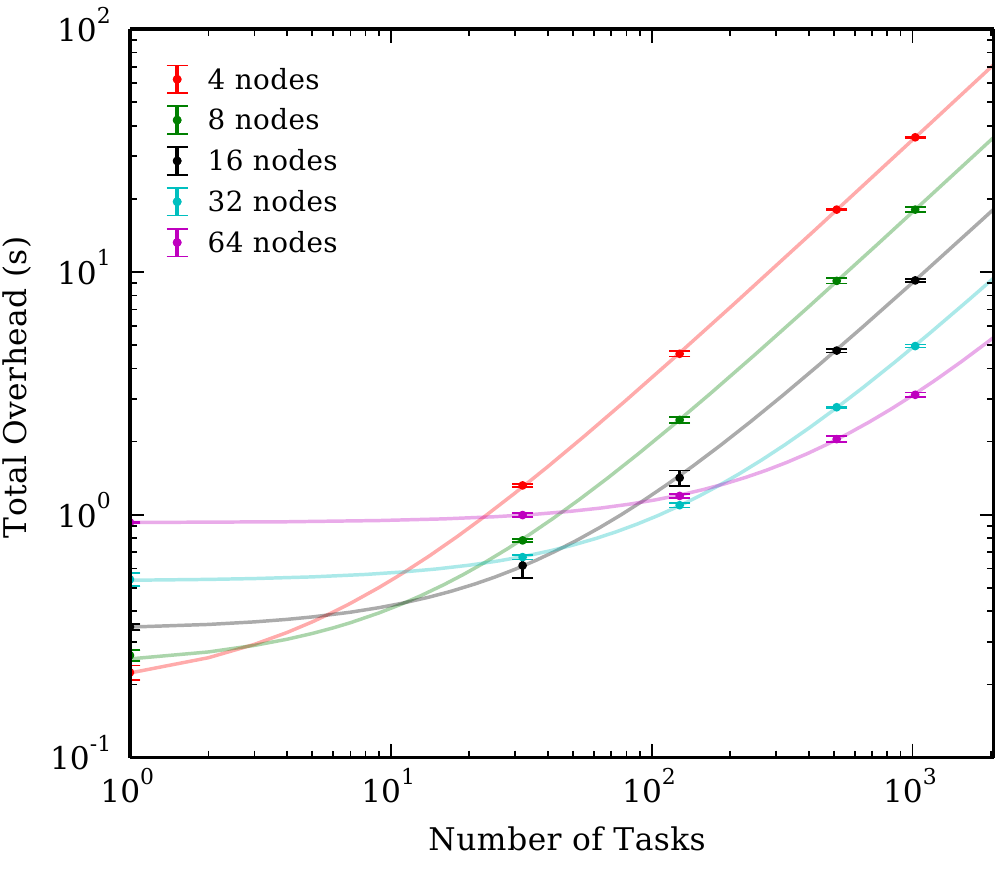}
\caption{\label{fig:task-overhead} The total overhead time as a function of the number of tasks
launched for various cluster sizes. The error bars show actual measurements using the EC2 hardware
configuration with four simultaneous tasks per node (i.e., $\nu=4$), while the curves show the
overhead estimated using Eq.~\ref{eq:overhead}, with appropriately-tuned time constants. The
overhead is roughly constant until the number of tasks exceeds the number of available execution
slots in the cluster (i.e., $N_{\rm tasks} > \nu N_{\rm nodes}$), it then grows in proportion to the
number of tasks. Note that larger clusters offer lower overhead for large numbers of tasks, but have
relatively high overhead for small numbers of tasks.}
\end{figure}

When data is stored in a distributed file system, such as HDFS, the overhead involved in
distributing the data among the various nodes is reduced---particularly when data locality is
exploited to minimize network traffic. However, there is still overhead; the tasks need to be
launched and the results need to be gathered. In Figure~\ref{fig:node-overhead} and
Figure~\ref{fig:task-overhead}, we present the overhead time, measured as described in
Section~\ref{section:methodology}, using EC2 clusters of various sizes, running 4 simultaneous tasks
per node.  We found that the scaling of the overhead time ($T_{\rm overhead}$) was well-modelled by
the relation,
\begin{equation}
 T_{\rm overhead} \approx \tau_0 + \tau_1 N_{\rm nodes} + \tau_2 N_{\rm cycles} \label{eq:overhead}
\end{equation}
where $N_{\rm nodes}$ is the number of worker nodes and $\tau_0$, $\tau_1$, and $\tau_2$ are time
constants, which depend upon the speed of the machines, the speed of the network, and details of the
software configuration.  The mean number of batch cycles per worker node is
\begin{equation}
 N_{\rm cycles} = \frac{N_{\rm tasks}}{\nu N_{\rm nodes}}, \label{eq:ncycles}
\end{equation}
where $N_{\rm tasks}$ is the total number of tasks submitted and $\nu$ is the number of simultaneous
tasks per worker node. For example, if the CPU(s) on each worker node of a cluster can execute 8
threads simultaneously and the user specifies that each task should be allocated 2 threads (by
setting \nbtt{threads_per_task=2}), then \nebulos\ will execute $8/2=4$ tasks on each node
simultaneously (i.e., $\nu = 4$). If the user then submitted 256 tasks to a cluster with 8 worker
nodes, then each worker node would execute $256/8 = 32$ tasks, on average, under the assumption that
all tasks require approximately the same amount of time to complete. Since each worker node can
process 4 tasks simultaneously, each node has to go through $32 / 4 = 8$ cycles. The time constant,
$\tau_2$, is the average overhead required for each cycle.

From Eq.~\ref{eq:overhead}, and Figures~\ref{fig:node-overhead} and \ref{fig:task-overhead}, we see
that the overhead time associated with launching a small number of tasks (i.e., $N_{\rm tasks}
\lesssim \nu N_{\rm nodes}$) increases as the number of nodes increases. However, when many tasks
(i.e., $N_{\rm tasks} > \nu N_{\rm nodes}$) are submitted to the cluster, larger clusters offer
lower overhead than smaller ones. We also see that the overhead is roughly constant until the
number of tasks exceeds the number of execution slots available in the cluster.

\subsubsection{Read speed}

\begin{figure}
\centering
\includegraphics[width=3.33in]{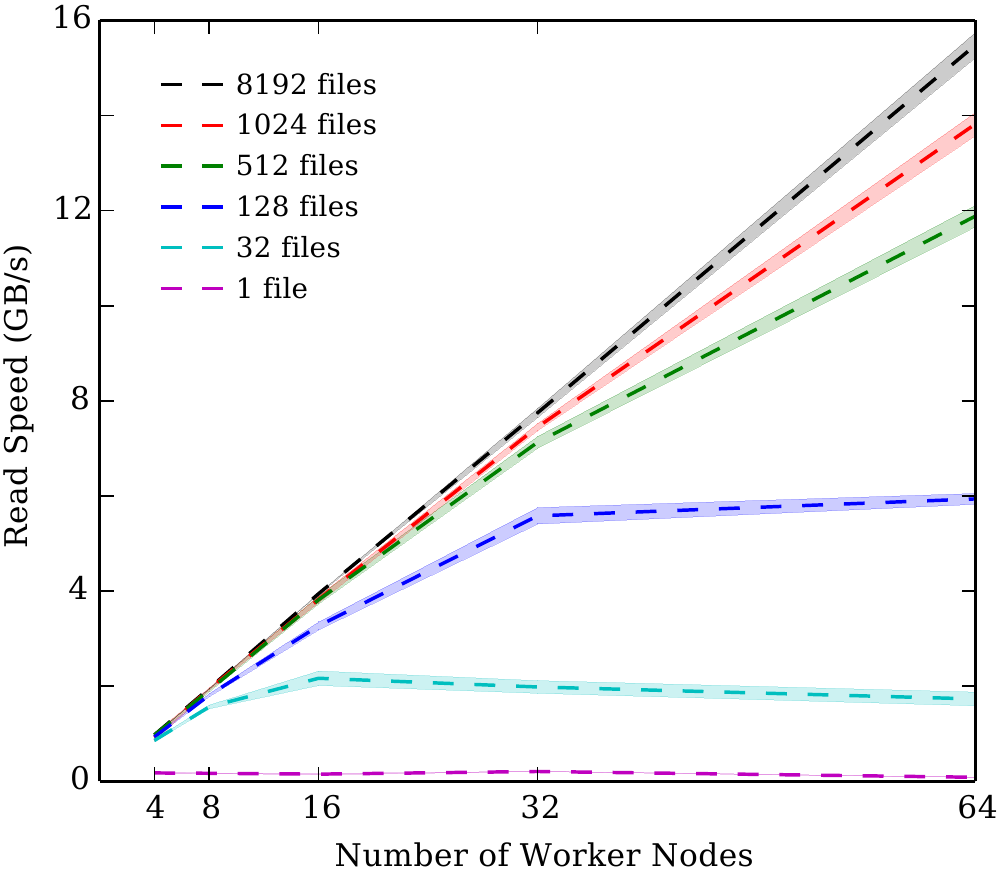}
\caption{\label{fig:node-scaling} The total read speed for datasets of various sizes, as a function
of cluster size when the EC2 system configuration is used. Dashed lines indicate the mean read
speed, while the shaded regions indicate the 1-$\sigma$ scatter. Note that the read speed is
proportional to the number of worker nodes, provided that a sufficient number of files is being
read. When the number of execution slots available in the system ($4 N_{\rm nodes}$, in this case),
exceeds the number of files in the dataset, the performance no longer increases; it tends to
decline slightly as the cluster grows. Also note that the total read speed generally exceeds the
total network throughput. For example, the network throughput for each node in the 64-node cluster
was roughly 110~$\rm MB~s^{-1}$ while the maximum sustained read speed achieved by each node was
approximately 250~$\rm MB~s^{-1}$, even when overhead was taken into account (the raw disk read
speed was approximately 420~$\rm MB~s^{-1}$).}
\end{figure}

In Figure~\ref{fig:node-scaling}, we present the net data reading rate for EC2 clusters of various
sizes. This demonstrates that the net throughput of the cluster scales linearly with the number of
worker nodes (up to at least 64 worker nodes), provided that the number of files being read is
sufficiently large. Note that \nebulos\ clusters tend to not reach their maximum-sustained reading
speed until each node contains $\sim\!\!300$ blocks, on average. Below this point, the dataset is
not distributed evenly enough among the nodes in order to reach optimal performance. More generally,
suppose the block replication factor is $\rho$ and we are interested in reading single-block files.
The dataset needs to contain $\gtrsim 300 N_{\rm nodes} / \rho$ files in order for the data to be
sufficiently well-distributed across the nodes of the cluster to reach the optimal reading speed.
From Figure~\ref{fig:node-scaling}, we also see that, when the number of files in the dataset is
smaller than the number of execution slots ($\nu N_{\rm nodes}$) in the cluster, adding more nodes
to the cluster tends to negatively impact the performance. This happens because the overhead
increases as the cluster grows, while the number of nodes that can read the data remains constant.
Note that decreasing $\nu$ can help to improve this situation somewhat, by forcing the tasks to be
distributed over more nodes.

It is possible to estimate the maximum sustained read speed, $R_{\max}$, achieved by \nebulos\ on a
cluster, if we make the following simplifying assumptions:

\begin{enumerate}
 \item The dataset consists of single-block files.
 \item Each task launched on the cluster reads only one file.
 \item All files are approximately the same size.
 \item The number of files in the dataset is much larger than the number of nodes in the cluster.
 \item The hardware of the worker nodes is
essentially uniform, so that no individual nodes are significantly faster or slower than the mean
and each worker node supports the same number of simultaneous threads.
\end{enumerate}

\noindent
Under these assumptions, the approximate size of the dataset is $DN_{\rm tasks}$, where $D$ is
the average size of the files being read. The total time required to read the dataset is then,
\begin{equation}
 T_{\rm total} \approx T_{\rm overhead} + \frac{DN_{\rm tasks}}{N_{\rm nodes}R_{\rm node}}
\label{eq:read-time}
\end{equation}
where $R_{\rm node}$ is the average sustained reading speed of a single worker node. The
maximum reading speed is,
\begin{equation}
  R_{\rm max} \approx \frac{DN_{\rm tasks}}{T_{\rm total}}.
\end{equation}
In the limit, $N_{\rm tasks} \gg N_{\rm nodes}$, this becomes
\begin{equation}
 R_{\rm max} \approx \frac{{N_{\rm nodes}}}{\tau_2 / \nu D + 1/R_{\rm node}}. \label{eq:max-read}
\end{equation}

Eq.~\ref{eq:max-read} agrees well with measurements of real systems and it can be used to guide
decisions about optimal usage and configuration settings. For instance, it shows us that reading
many small files (small $D$) negatively impacts the read speed. This happens because the
task-launching overhead is more significant when many small files are being read, compared to a
smaller number of larger files. Conversely, increasing the HDFS block size on large files will
increase the read performance. Launching more simultaneous tasks per node ($\nu$) will also increase
the performance, provided that the tasks are I/O bound and not computationally intensive. In
general, increasing the speed of the data storage media on the nodes ($R_{\rm node}$) will increase
the performance. However, due to overhead, the speed-up factor realized by the cluster is not
proportional to the node speed-up factor.

Once the number of worker nodes becomes sufficiently large, the computational speed of the Mesos
Master and HDFS NameNode host machines (or the speed of their network interfaces) will prevent the
scaling from being purely proportional to the number of worker nodes. For instance, as the number of
worker nodes increases, the value of $\tau_2$ will eventually increase, due to increased network
traffic at the Mesos Master host. This likely only becomes an issue in clusters with more than
$\sim$1,000 nodes with current hardware. Note that Mesos and HDFS both scale well to at least
$\sim$10,000 nodes and we have no reason to suspect that the \nebulos\ Application Framework would
hinder the scalability of the system.

\subsection{Multi-block files}

\begin{figure}
\centering
\includegraphics[width=3.33in]{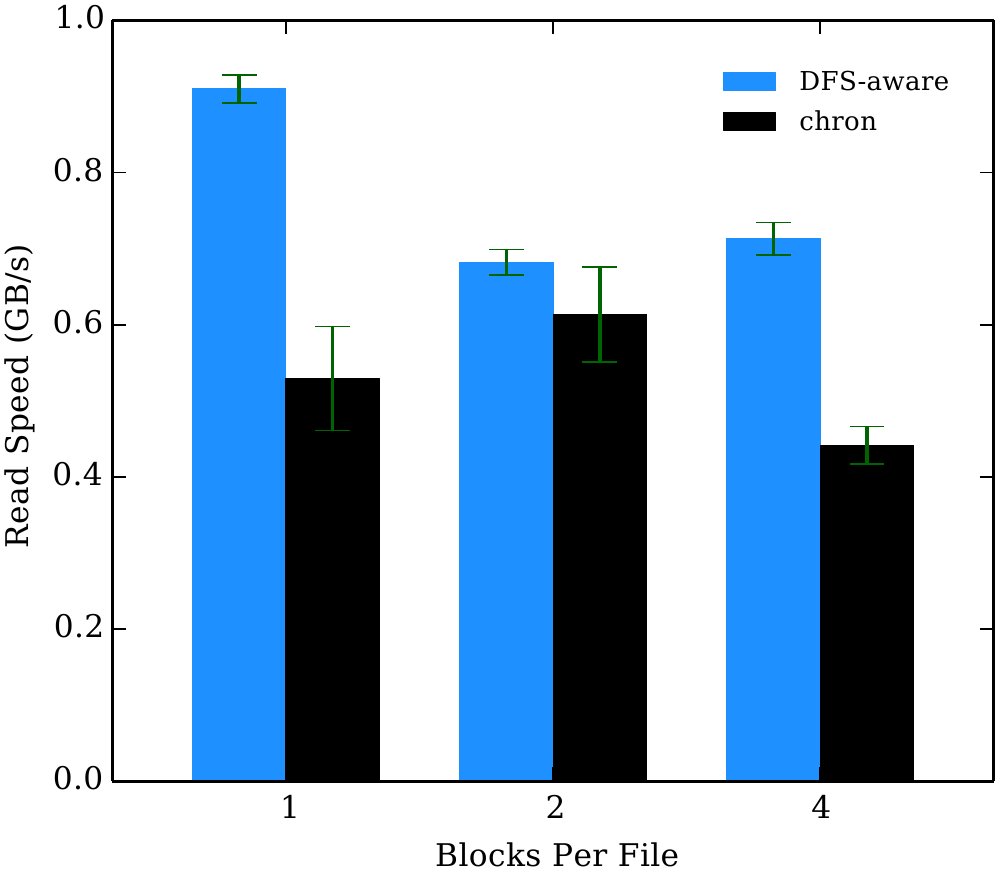}
\caption{\label{fig:block-scaling} The read speed for three datasets of the same size, but
distributed over different numbers of files. The first dataset consisted of 1000 single-block flies,
the second consisted of 500 two-block files, and the third consisted of 250 four-block files. All
measurements were made on the MicroBlade cluster. Note that the DFS-aware scheduler consistently
outperformed the chronological scheduler and the data set consisting of single-block files was read
more efficiently than the multi-block datasets.}
\end{figure}

In the analysis above, we have only examined the speed with which files consisting of a single HDFS
block can be read. In this section, we investigate \nebulos' performance when reading multi-block
files on the MicroBlade cluster. Recall that this cluster only had 8 worker nodes. We began by
reading a dataset containing 1000 single-block galaxy simulation snapshot files, with a total size
of 110.8~GB. We then combined these files into a set of 500 two-block files of the same total size
and then combined the 500 files into a set of 250 four-block files. The resulting read speeds
achieved by the DFS-aware and chronological schedulers are presented in
Figure~\ref{fig:block-scaling}.

Keep in mind that the HDFS block size can be specified on a per-block basis. Thus, with a
small amount of effort, the user can typically ensure that even multi-gigabyte files only consist of
a single HDFS block. Figure~\ref{fig:block-scaling} demonstrates why a user may want to take this
extra step; the read speed of multi-block files tends to be slower than that of single-block files.
The cause of this behaviour is straightforward; when files are broken into blocks, they are
distributed across different nodes, so a portion of the data typically needs to be transferred over
the network. Whenever possible, the DFS-aware scheduler will try to launch tasks on machines that
contain all of the blocks associated with a file. However, this is not always possible. Typically,
the DFS-aware scheduler will assign tasks to nodes which contain only a portion of a file's
constituent blocks. This still results in better performance than the chronological scheduler,
which relies heavily upon network transfers.

It should also note that, as the number of blocks in a file increases, the chance of multiple 
blocks being stored on the same node increases. The number of nodes that can read a portion of the 
data from local storage also increases. This effect is more significant for small clusters than 
large ones.

In our analysis, there was an additional effect that partially compensated for the increased
reliance upon the network as the number of blocks increased. Since the number of files decreased as
the number of blocks increased, there was less task-launching overhead for the multi-block files.
This reduced overhead can be seen from Eqs.~\ref{eq:overhead} and \ref{eq:ncycles}; the number of 
cycles decreased when the number of tasks decreased, which lead to lower overhead.

\section{Discussion}
\label{section:discussion}

\nebulos\ combines the strengths of Big Data tools and classic batch processors. Like a classic 
batch processor, \nebulos\ can launch arbitrary software a cluster. In contrast, most software used 
by popular Big Data frameworks must be made aware of the framework in some way. The tasks launched 
by \nebulos\ can operate on arbitrary data formats with little or no extra effort from the user, 
whereas most Big Data tools require extra effort for each specific data format that is used. Unlike 
classic batch schedulers, \nebulos\ tasks can be scheduled in a data locality-aware manner in order 
to improve data throughput. New tasks can be incrementally added to a batch job, results can be 
accessed programmatically while the job is running, and multiple batch jobs can be used in a single 
analysis routine---either in parallel or sequentially. Additionally, \nebulos\ provides an automated 
way to monitor the behaviour of each task and take actions, based upon the observed behaviour. This 
is a feature that no other Big Data tool currently offers.

Since \nebulos\ is based upon industry-standard tools, it benefits from the efforts of a large
community of engineers. This also means that the platform is compatible with many existing tools.
For instance, Apache Hadoop MapReduce, \citetalias{Spark}, \citetalias{Hama}, \citetalias{Storm},
\citetalias{Torque}, and \citetalias{MPICH} can all run on the \nebulos\ platform alongside the
\nebulos\ application framework because all of these tools are compatible with the Apache Mesos
kernel. It is even possible to use the \nebulos\ application framework and Apache Spark together in
the same Python script.

When locality-aware distributed file systems, like HDFS were designed, most datacenters relied on
1~Gbps networks. Modern datacenters use 40~Gbps and 100~Gbps networks. These high-performance
networks have alleviated the network bottleneck that motivated the design of locality-aware file
systems. At the same time, though, the speed and cost of solid state hard drives has improved, which
means that locality-aware frameworks, like \nebulos, are still able to outperform systems that do
not make use of data-locality. Locality-aware systems can also be built in a more cost-effective 
manner because less expensive networking hardware can be used without significantly reducing the 
I/O performance.

\subsubsection*{\nebulos\ as a cloud service}

Cloud computing services allow users to pay for computing resources as needed, rather than building
their own clusters. These services are especially useful when computing resources are only needed
for short-term projects. Installing \nebulos\ on cloud services is straightforward, once the
constituent parts have been compiled. We have developed a system image and installation scripts for
Amazon Web Services, which allows users to build a \nebulos\ cluster in Amazon's Elastic Compute
Cloud in less than 15 minutes. Since the \nebulos\ source is openly available, users are free to
create their own installers for other cloud services.

\subsubsection*{Future work}

Although it is useful to treat HDFS as a local file system, as is done in \nebulos, the user must be
aware of certain complications that can arise. In particular, since FUSE-DFS does not support
appending data to files and HDFS does not allow data to be modified once it is written, programs
that depend upon the ability to append and modify data do not work properly without user
intervention. This can cause errors that are difficult to diagnose. Also, storing a large number of
small files (much smaller than the HDFS block size) unnecessarily burdens the HDFS NameNode. This
problem can be alleviated by combining small files into larger files, but this requires extra
effort. Dealing with very large files (requiring tens of blocks) is also inefficient because the
entire file must be read by each task that uses it. When possible, splitting large files into
smaller, independent pieces, can help to resolve this problem. However, this also requires extra
effort by the user. For maximal throughput, programs sometimes need to directly access the HDFS
using the HDFS library. FUSE-DFS also sometimes fails to write large files to the HDFS in a timely
manner; the user has to instruct the system to synchronize if a large output file will be read soon
after it is written.

When implementing an algorithm that requires the same data to be accessed repeatedly by subsequent
tasks, it is beneficial to cache the repeatedly-used data by saving it on a local hard drive or in a
RAM disk. This improves performance because there is no need to repeatedly read the same files from
the HDFS. In its current form, \nebulos\ does not provide an optimal method of caching data. The
user can launch tasks which store data in a specific location on the local machine, however the user
is also responsible for deleting such files when they are no longer needed. Furthermore, there is no
easy way to ensure that subsequent tasks are launched on machines containing cached data.

\nebulos\ offers no automatic way to facilitate inter-task communication. In order for tasks to
communicate, the user must either save files to the HDFS or include network communication
capabilities in the task software. This requires extra thought and effort on the part of the user.

We plan to make the \nebulos\ application framework even easier to use by providing automated
methods for handling additional usage scenarios. We will also work to address the limitations of the
\nebulos\ application framework, discussed above. In particular, we plan to add an easy, automated
method for persistently storing certain data on local disks (including RAM disks) for use by
subsequent tasks. Cached data will automatically be cleaned up so that manual intervention is
unnecessary. We plan to provide an automated means of enabling inter-task communication so that a
broader variety of algorithms can be implemented easily. The \nebulos\ application framework will
rely less heavily upon FUSE-DFS and there will be facilities for automatically improving the
performance of tasks that use large numbers of small files as well as some tasks that make use of
very large files. It may eventually be possible to completely replace HDFS with a modified version
of CephFS \citep{Ceph-paper} that allows client software to query the physical
location of data (CephFS does not currently provide this feature).

\section{Acknowledgements}

This research was funded by a Big Data seed grant from the Vice chancellor for research and
development, UC Riverside.

\bibliographystyle{mnras}
\bibliography{references.bib}

\appendix

\section{Fault tolerance}
\label{appendix1}
In this appendix, we describe the fault tolerance features of Mesos and HDFS in greater detail.

\subsection*{Mesos}

If the Mesos Master daemon detects that a slave has become unreachable (for instance, due to
hardware failure), it notifies the relevant framework schedulers that the tasks running on the
unreachable machine have been lost. The schedulers can then assign the lost tasks to other hosts.
Thus, the system seamlessly handles node failures. Additionally, it is possible to configure
redundant Mesos Master daemons, so that the system is resilient to master node failure.

\subsection*{HDFS}

The HDFS DataNode daemon sends status messages, called ``heartbeats,'' to the NameNode at regular
intervals to inform the NameNode that it is still alive and reachable. If the NameNode stops
receiving heartbeats from a particular DataNode, that DataNode is assumed to be dead. The NameNode
then instructs the remaining DataNodes to make additional copies of the data that was stored on the
dead node so that the required replication factor of each block is maintained. Additionally, each
time a DataNode reads a block of data, it computes a checksum. If the checksum does not match the
original checksum, stored in the block's associated metadata file, the NameNode is informed that the
block has been corrupted. A new copy is then created from the uncorrupted copies on other machines.
As in the case of the Mesos Master daemon, it is possible to configure redundant, backup NameNodes
so that the file system remains intact when the machine hosting the primary NameNode experiences a
failure.

\section{Summary of existing tools}
\label{appendix2}

In this appendix, we briefly summarize the features of several popular tools: Apache Hadoop
MapReduce, Apache Hama, Apache Spark, Slurm, and TORQUE. We also comment on aspects of these
tools that limit their usefulness for large-scale scientific data analysis.

\subsection*{\rm \bf Hadoop MapReduce}

The Hadoop MapReduce framework is an implementation of the MapReduce programming model. The user
provides a mapper function and a reducer function, typically in the form of Java class methods. The
user then provides a set of key-value pairs (for instance, file names and file contents) for the
framework to operate upon. The mapper transforms the initial set of key-value pairs into a second
set of key-value pairs. The intermediate key-value pairs are then globally sorted by key and
transformed into a third set of key-value pairs by the reducer.

\subsubsection*{Usability notes}

The mapper and reducer almost always need to be designed with Hadoop in mind. A feature called
Hadoop Streaming makes it possible to use pre-existing executable files as the mapper and reducer.
However, the executables need to be able to read and write key-value pairs via the standard input
and output streams. In many situations, using pre-existing software with Streaming requires the
software to be invoked from a script which formats the input and output streams appropriately. In
other situations, the mapper and reducer programs have to be modified in order to work properly.

Handling non-trivial data formats efficiently requires special care. If the data format being used
with MapReduce is more complicated than a text file that can be split into single-line records, then
a custom file reader must be defined in order for the MapReduce framework to properly read the data.
Reading binary files produced by scientific instruments or simulations is even more cumbersome than
reading formatted text.

Using Hadoop MapReduce with languages other than Java requires a bit more work, since the framework
is primarily intended to be used by Java programmers. In order to use Hadoop MapReduce with
languages other than Java, one can use Hadoop Pipes, which makes it possible to write mappers and
reducers in C++. The C++ code can be extended to make use of other languages, such as Python.

\subsection*{\rm \bf Hama}

\citetalias{Hama} is a framework for Big Data analytics which uses the Bulk Synchronous Parallel
(BSP) computing model \citep{BSP} in which a distributed computation proceeds in a series of
super-steps consisting of three stages: (1) concurrent, independent computation on worker nodes, (2)
communication among processes running on worker nodes, and finally (3) barrier synchronization.

Individual processes stay alive for multiple super-steps. Thus, data can easily be stored in RAM
between steps. This allows Hama to perform very well on iterative computations that repeatedly
access the same data. Hama can outperform Hadoop MapReduce by two orders of magnitude on such tasks.

\subsubsection*{Usability notes}

Like Apache MapReduce, Hama is primarily intended to be used with Java, but it is possible to write
programs in C++, using Hama Pipes. Hama handles data formats in exactly the same way as Hadoop
MapReduce. Thus, working with raw scientific data is not straightforward in most cases.

\subsection*{\rm \bf Spark}

\citetalias{Spark} is framework based upon the concept of Resilient Distributed Datasets (RDDs)
\citep{RDD}. As the name suggests, RDDs are distributed across a cluster of machines. For added
performance, the contents of an RDD can be stored in memory. Their resiliency lies to the fact that
only the initial content of the RDDs and transformations performed on them need to be stored in a
distributed file system; the memory-resident version of an RDD can be automatically recreated upon
node failure by repeating transformations on the initial data.

Compared with Hadoop MapReduce, Spark offers more flexibility. There is no need to use a particular
programming model; it is possible to implement MapReduce, BSP, and other models with Spark. Spark is
also more interactive. Once an RDD is created, operations can easily be performed on the data by
issuing commands from an interpreter. Spark works natively with four programming languages: Scala,
Java, Python, and R. It also provides a module allowing queries to be written in SQL.

\subsubsection*{Usability notes}

Any executable file can be invoked with the RDD \nbtt{pipe()} transformation, which sends data to
the executable via a Unix pipe and then stores the standard output of the executable in an RDD.
However, in order to be useful, the program must be able to read data from its input stream and send
output data to the standard output stream. Programs that do not behave in this way need to first be
modified in order to be compatible with Spark.

Using data that is more complicated than plain text requires the user to define a file format
reader. Thus, working directly with raw scientific data formats is not straightforward.





\subsection*{\rm \bf TORQUE}

\citetalias{Torque} is a distributed resource manager, designed for submitting and managing batch
jobs on a cluster. With TORQUE, it is possible to launch arbitrary programs and operate on arbitrary
data. Batch jobs are launched by first writing a batch submission script and then submitting the
script to the scheduler. It is possible to monitor the status of a batch job and individual tasks
while the job is running.

\subsubsection*{Usability notes}

Unlike the Big Data tools, discussed above, TORQUE is not aware of data placement. Thus, data
throughput depends heavily upon the speed of the network. It is also not straightforward to create
data analysis routines consisting of multiple batch jobs. TORQUE is primarily intended for manually
launching batch jobs, rather than creating jobs programmatically.

\subsection*{\rm \bf Slurm}

The Simple Linux Utility for Resource Management \citepalias{Slurm} is an open source,
fault-tolerant, and highly scalable cluster management and job scheduling system for Linux clusters.
It is used by some of the world's fastest supercomputers. The architecture of Slurm is quite similar
to the architecture of Mesos + \nebulos\ application framework.

\subsubsection*{Usability notes}

Like \nebulos, Slurm allows jobs to be launched via a script or interactively. It can also be used
to launch arbitrary software on a cluster. Slurm offers several different schedulers, but none are
optimized to take advantage of data locality. However, it would likely be fairly straightforward to
add a new scheduler to Slurm that could take advantage of data locality awareness.

\subsection*{\rm \bf CephFS}

The Ceph Distributed File system \citep[\citetalias{Ceph}, ][]{Ceph-paper} is an open source
implementation of the reliable autonomic distributed object store architecture \citep[RADOS,
][]{RADOS}. It is a POSIX-compliant, high-performance, distributed file system with automatic data
replication for fault-tolerance. CephFS is able to scale to multiple exabytes because it distributes
its metadata server across many machines (whereas the metadata in HDFS is stored on a single
NameNode host). It also offers many performance enhancements, such as automatically increasing the
replication factor of files that are frequently accessed (HDFS requires this to be done manually).
CephFS is implemented as a Linux kernel module, so it runs in kernel space, just as a native
file system, rather than in user space.

\subsubsection*{Usability notes}

CephFS does not provide a straightforward way for client software to query the physical location of
data. Thus, it relies heavily upon network transfers.

\bsp	
\label{lastpage}
\end{document}